\documentstyle[prl,aps,graphicx,multicol]{revtex}
\renewcommand{\narrowtext}{\begin{multicols}{2} \global\columnwidth20.5pc}
\renewcommand{\widetext}{\end{multicols} \global\columnwidth42.5pc}

\begin{document}
\title{Electrical Manipulation of Nanomagnets}
\author{ L. Y. Gorelik$^{1,2}$, R. I. Shekhter$^{1,2}$, V.Vinokur$^2$, D.Feldman$^2$,
V.Kozub$^{2,3}$
 and M. Jonson$^1$}
\address{$^1$Department of Applied Physics, Chalmers University of
Technology, SE-412 96 G\&quot;oteborg, Sweden.\\
$^2$ Material Science Devision Argonne National Laboratory, USA\\
$^3$Division of Solid State Physics, Ioffe Institute of the Russian Academy of Sciences, St.
  Petersburg 194021, Russia.}

\date{\today}
\maketitle

\begin{abstract}We demonstrate a possibility to manipulate the magnetic
coupling between two nanomagnets with a help of ac electric field.
In the scheme suggested the magnetic coupling in question is
mediated by a magnetic particle contacting with both of the
nanomagnets through the tunnel barriers. The electric field
providing a successive suppression of the barriers leads to
pumping of magnetization through the mediating particle. Time
dependent dynamics of the particle magnetization allows to to
switch between ferro- and antiferromagnetic couplings.
\end{abstract}

\vspace*{0.1in}
\pacs{PACS number: ????}
\narrowtext

The sensitivity of electron transport to the spin degree of
freedom brings new possibilities for implementing device functions
in electronics. As a result the field of spintronics is developing
rapidly. The giant magneto-resistance \cite{GMR} is a striking
example of an effect of spin dependent transport that has already
found important applications in computer hardware. More
fundamental ideas for using spin in order to realize devices that
can store and process quantum information are now under intensive
discussion in the literature \cite{Kan,Kikkawa}.

Manipulation of the electron spin is only possible if one is able
to control the magnetization of the magnetic materials that are
necessary elements of any spintronics device. In nanoscale devices
a fundamental obstacle to achieve the required level of control
comes from the fact that the magnetic fields used to control the
magnetization cannot be localized on the nanometer length scale.
This is in sharp contrast to the electric fields used in modern
nanoelectronics based on the Single-Electron devices
\cite{Liharev}. The problem of  selective control of the
magnetization has therefore become crucial for functioning of the
nanoscale spintronics devices. A use of electric rather than
magnetic fields to manipulate nanomagnets could, if it works, be a
way out of this ``nonlocality trap". A natural way to realize such
a control is to make use of the indirect exchange interaction
between nanomagnets induced by conducting electrons. Indeed, in
the hybrid structures where ferromagnetic layers are separated by
normal metals the indirect exchange can be controlled electrically
by affecting the wave functions of electrons mediating the
exchange \cite{Shwabe, Bader, Kozub}. In this case the transfer of
spin polarization between the ferromagnetic layers is controlled
by an interference pattern produced by different electronic waves
and therefore crucially affected by any kind of structural
material disorder. Since the latter is obviously dependent on the
atomic scale details of interface geometry, the phenomenon becomes
very sensitive to fluctuations and noise in the system.

The main idea of the present paper is to explore a new possibility
of magnetic coupling where a magnetization is transferred through
some "time domains" rather than through the spatial domains. Such
a possibility occurs if magnetic coupling between two nanomagnets
is mediated by a small magnetic particle ("mediator").
Accumulation of magnetization transferred from one nanomagnet to
another in an "intermediate state" on mediator enables one to
realize a delay line with the possibility to control a magnitude
and orientation of transferred magnetization. Electrical
manipulation of nanomagnets becomes possible if exchange
interaction, which is essentially of electrostatic origin, is
employed.

A sketch of the structure to be considered is presented in Fig. 1.
\begin{figure}
\centerline{\includegraphics[width=5cm]{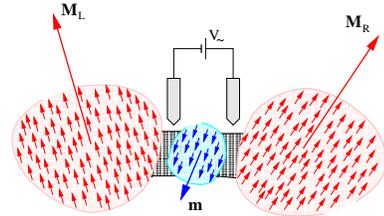}}
\vspace*{0.5cm} \caption{Schematic diagram of the system discussed
in the text. Single domain magnetic grains with magnetic moments
${\bf M}_L$ and ${\bf M}_R$ are coupled via the magnetic cluster
with magnetic moment ${\bf m}$, the latter being separated from
the grains by insulating layers. The gate electrodes induce an ac
electric field, concentrated in the insulating regions. This field
controlling the heights of the tunnel barriers affects the
exchange magnetic coupling between different components of the
system.} \label{f_fig1}
\end{figure}
The figure shows two single-domain nanomagnets with magnetic
moments $\mathbf{M}_{L}$ and $\mathbf{M}_{R}$. They are both
coupled by the direct exchange interaction spreading through the
corresponding tunnel barriers  to a magnetic cluster or magnetic
molecule with the magnetic moment $\mathbf{m}$. So the
cluster/molecule acts as a magnetic weak link between the magnets.
An indirect exchange interaction between the two nanomagnets is
mediated by the cluster/molecule which acts as a magnetic weak
link between the magnets. Note that the exchange coupling between
mediator and the magnetic leads is controlled by the heights of
the tunnel barriers that separate the electronic states of the
nanomagnets and the cluster. We will show that a periodic electric
field applied to the tunnel barriers (inducing a time-dependent
exchange coupling) can transform the character of the mediated
exchange between the nanomagnets from being ferromagnetic to
antiferromagnetic one. We will assume that the exchange coupling
between mediator and leads has a time dependence that corresponds
to a sequential coupling of the mediator to first one of the
magnetic leads and then to another one, in a periodically
repeating pattern (the heights of the tunnel barriers oscillate
with a phase shift of $\pi$). In this case three stages of the
mediated coupling between the leads can be distinguished: 1)
polarization of the mediator by one of the leads (while the
mediator is essentially decoupled from the other one); 2) the
internal dynamics of the free mediator  (this occurs when the
mediator is decoupled from both leads); 3)transfer of the induced
magnetic polarization from the mediator to the second lead (while
decoupled from the first one).
For simplicity we will omit this step, assuming that there is no
nontrivial dynamics of the mediator spin when decoupled from the
leads. Under these conditions the time evolution of $\mathbf{m}$
can be thought of as being due to a sequence of "scattering
events". A single "scattering" results in a change of the mediator
magnetic moment by the value $\Delta\mathbf{m}$. On the other
hand, due to the conservation of magnetic momentum, a magnetic
moment change takes place also in the lead after the "scattering"
event. Therefore, one can look at the process as being the
mediator-assisted flow of magnetic polarization between the leads.
This flow, giving rise to a synchronized evolution of the
magnetization in the leads, establishes an effective coupling
between them.

 Since
$M >> m$ the dynamics of the magnetization in the leads is much
slower than the dynamics of the magnetic moment (spin) of the
mediator. When considering the dynamics of the mediator
magnetization, one can therefore to a first approximation neglect
the variation of $\mathbf{M}$ altogether. Thus the time-dependent
exchange coupling of the mediator to the leads will result in an
effective periodically oscillating magnetic field acting on the
magnetic moment of the mediator.
As we will prove below, any weak relaxation  will bring the
mediator magnetization $ {\mathbf m}(t)$ into a periodic regime
for which $ {\mathbf m}(t) = {\mathbf m}(t+2T)$. In this regime
the magnetic moment of the mediator changes from a value ${\mathbf
m}_{1}$ to another value ${\mathbf m}_{2}$ during a first
half-period when the mediator cluster is coupled to the left lead,
and vice versa (from ${\mathbf m}_{2}$ to ${\mathbf m}_{1}$)
during a second half-period when it is coupled to the right lead.
While being coupling to a lead, the mediator being affected by an
effective magnetic field with fixed direction and its moment
rotates around an axis parallel to the magnetization of the lead
(see Fig.2).

\begin{figure}
\centerline{ \includegraphics[width=5cm]{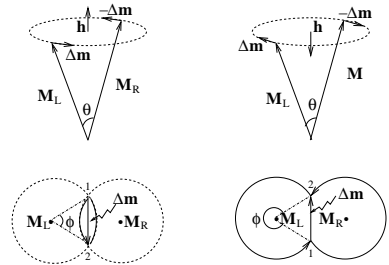} }
\vspace*{0.5cm} \caption{Schematic diagram demonstrating the
periodic
The bottom part represents periodic dynamics of the projection of
a mediator magnetization on the plane perpendicular to the vector
${\bf M}_L +{\bf M}_R$. Points L and R represent the axes aligned
vectors ${\bf M}_L$ and ${\bf M}_R$ correspondingly. If molecular
cluster is coupled to one of the nanomagnets, its magnetic moment
rotates counter-clockwise around the axes L or R (depending on
what nanomagnet it is subject to). The circles schematically
represent the trajectories which are traced out by the end of the
vector $\bf{m}$. The angle of rotation $\phi$ depends on the
length of time interval during which the mediator is coupled to
the nanomagnets  and on the intensity of exchange coupling. The
mediator magnetization evolution is matched to be the oscillations
between points 1 and 2 after each half-period. Magnetic moment
$\Delta {\bf m}$ is transferred from one nanomagnet to another one
during the period setting them into rotational motion around axes
${\bf M}_L +{\bf M}_R$ (upper part of the figure).} \label{f_fig1}
\end{figure}
The total angle of rotation $\phi_{\alpha} =
g\bar{J}_{\alpha}M_{\alpha}T$, after the mediator has been
magnetically coupled to the lead for a certain amount of time
during one contact, depends on the average exchange coupling
strength $\bar{J_{\alpha}}$ and the effective coupling time T ($g
= 2\mu/\hbar$, $\mu$ is the Bohr magneton). One finds that in the
symmetric case ($\phi_{L} = \phi_{R} = \phi$) that we will
consider in this report the vector $\Delta\mathbf{m}$ is
perpendicular to the plane spanned by $\mathbf{M}_{L}$ and
$\mathbf{M}_{R}$) (below denoted the $xy$-plane). The flow of
polarized magnetization will result in a rotation of
$\mathbf{M}_{\alpha}$ around an axis parallel to the vector
$\mathbf{M}_{L}+\mathbf{M}_{R}$ ( below x-axes).
It can be described as an effect of some magnetic field
$\mathbf{h}$ directed along that axes (see Fig.2). Relaxation
processes, that are inevitably present, will tend to align the
magnetization of the lead along this field. Let us suppose that
the rotation angle $\phi =\phi_{0}$ is much smaller than $2\pi$.
Under this condition the vectors $ {\mathbf m}_{1,2}$ will be
aligned nearly along the bisector of the angle between ${\mathbf
M}_{L}$ and ${\mathbf M}_{R}$ and therefore $\mathbf{h}$ will be
directed along the vector $\mathbf{M}_{L}+\mathbf{M}_{R}$. In such
a case the magnetic moments of the leads, since they tend to be
aligned along the effective magnetic field, will obey a
ferromagnetic order ($\theta =0 $). Now let us
assume that the rotation angle is $2\pi - \phi_{0} > \pi$. One
finds, that if a rotation by an angle $\phi_{0}$ around some axis
gives rise to a change of the magnetic moment from ${\mathbf
m}_{2}$ to ${\mathbf m}_{1}$, the rotation around the same axis by
the angle $2\pi - \phi_{0}$ will transform ${\mathbf m}_{1}$ into
${\mathbf m}_{2}$. Therefore, the periodic evolution of ${\mathbf
m}(t)$ will be established in the way that during the first half
period (when the mediator is coupled to the right lead) its moment
changes from ${\mathbf m}_{1}$ to ${\mathbf m}_{2}$ and vice versa
during the second half period. So we will have the same magnetic
flow, but in the opposite direction.
Thus one concludes that the effective magnetic fields at
$\phi_{0}$ and at $2\pi - \phi_{0}$ will be pointing in opposite
directions. Consequently, at $\phi = 2\pi - \phi_{0}$ the $
\mathbf{h}$ should be anti-parallel to the vector ${\mathbf
M}_{L}+{\mathbf M}_{R}$ making the ferromagnetic ordering
unstable. Below we will show that if $ \phi
> \pi$ the system
exhibits an antiferromagnetic ordering. Therefore, by tuning the
rotation angle $\phi$ --- which depends on the amplitude and
frequency of the alternating electric field
--- one can switch from ferromagnetic to antiferromagnetic coupling between
the magnetizations in the leads. For a quantitative discussion of
the phenomena outlined above, we will use Landau-Lifshits
equations:
\begin{eqnarray}\label{LLm}
\frac{1}{g}\frac{d{\mathbf m}}{dt} =(\frac{\partial
W}{\partial{\mathbf m}}\times{\mathbf m}) +
\frac{\beta}{|m|}({\mathbf m}\times( {\mathbf
m}\times\frac{\partial
W}{\partial{\mathbf m}}))\\
\frac{1}{g}\frac{d{\mathbf M}_{\alpha}}{dt} = (\frac{\partial
W}{\partial{\mathbf M}_{\alpha}}\times{\mathbf M}_{\alpha}) +
\frac{\beta}{M}({\mathbf M}_{\alpha}\times( {\mathbf
M}_{\alpha}\times\frac{\partial W}{\partial{\mathbf
M}_{\alpha}}))\nonumber
\end{eqnarray}

Here $M = |\mathbf{M}_{\alpha}|$, $m = |\mathbf{m}|$ and magnetic
energy of system $W$ has a form:
\begin{equation}\label{W}
W = -\sum_{\alpha =L,R}J_{\alpha}(\mathbf{M}_{\alpha}{\mathbf{m}})
\end{equation}
where $J_{\alpha}(t)$ describes a periodic (with the period $2T$)
time-dependent exchange coupling between mediator and magnetic
leads. In this paper we take $J_{L,R}(t) = \bar{J}(1 \pm
\alpha(t))/2$ with $\alpha(t) = sign(\sin\pi t/T)$. The second
terms in equations (1) describe the relaxation with relative
characteristic frequency
$\beta$.
In what follows we will assume  $\beta \ll 1$, (according to
literature \cite{Slanch} $\beta$ varies from 0.5 to 0.005
depending on magnetic material). IN this case the dissipation only
slightly affects the magnetization dynamics and non-trivial
regimes  can be expected. If $M \gg m $, the dynamics of molecular
spin is much faster then the dynamics of leads magnetization, and
one can use adiabatic approximation to analyze the behavior of the
system. To do this we will calculate $\mathbf{m}(t)$ under
assumption that the magnetization of the leads is fixed and then
substitute it into the equation (2). Then averaging over the fast
oscillation one obtains the following equation for
$\mathbf{M}^{\alpha}$:
\begin{equation}\label{Mef}
{\frac{1}{g}\frac{d\mathbf{M_{\alpha}}}{dt} =
(\bar{\mathbf{h}}^{\alpha}\times\mathbf{M}_{\alpha})} +
\frac{\beta}{
M}(\mathbf{M}_{\alpha}\times(\mathbf{M}_{\alpha}\times
\bar{\mathbf{h}}^{\alpha}))
\end{equation}
where the effective magnetic fields $\bar{\mathbf{h}}^{\alpha}$
are given by the relation $\bar{{\mathbf h}}^{\alpha} =
(2T)^{-1}\int^{2T}_{0} dtJ_{\alpha}(t)\mathbf{m}(t)$. Therefore
the dynamics of the leads magnetization is controlled
 by average spin polarization of the mediator when it is
coupled to the lead. Integrating equation (1) over period, we
obtain ${\mathbf m}(2T) - {\mathbf m}(0) = gT\{({\mathbf
M}_{L}\times\bar{{\mathbf h}}^{L}) + ({\mathbf
M}_{R}\times\bar{{\mathbf h}}^{R})\}$. It means that in the case
of periodic evolution (${\mathbf m}(2T) = {\mathbf m}(0)$)
 the average fields $\bar{\mathbf{h}}^{L,R}$
 obey the relations $({\mathbf M}_{L}\times\bar{{\mathbf h}}^{L}) =
 -({\mathbf M}^{R}\times\bar{{\mathbf h}}^{R})$. Taken scalar
 product of this relation with $\mathbf{M}_{\alpha}$ one can
 easily find
 that the projection of $\bar{\mathbf{h}}^{\alpha}$ on the
axis perpendicular to $({\mathbf M}^{L},{\mathbf M}^{R})$-plane
(below xy-plane) is equal to zero in
 the periodic regime.
 As a result, $\bar{\mathbf{h}}^{\alpha}$ may be presented as a linear
 combination of magnetizations
 $A{\mathbf M}^{\alpha} +
 \bar{L}{\mathbf M}^{\beta}$, where coefficient $\bar{L}$ is some
function of the angle
 $\theta$ between
 the vectors ${\mathbf M}^{L}$ and ${\mathbf M}^{R}$.
 One can represent the magnetic fields through effective inter-leads
interaction energy $ \mathcal{W}$:
 $$ \mathbf{h}^{\alpha} = -\frac{\delta
\mathcal{W}}{\delta\mathbf{M}^{\alpha}}$$
 The structure of the effective potential $\mathcal{W}$
 controls the type
(ferromagnetic or antiferromagnetic)
 of the interaction between
 the nano-magnets.
 Making use of
 the fact that $\mathcal{W}$  depends only
on the angle $\theta$, and consequently can be represented as a
function of scalar product $({\mathbf M}^{L}\cdot{\mathbf
M}^{R})$, one can prove
 the following
 relations:$({\mathbf e}_{z}\cdot({\mathbf M}^{L}\times{\mathbf h}^{L}))
 = - ({\mathbf e}_{z}\cdot({\mathbf M}^{R}\times{\mathbf h}^{R})) =
 \partial {\mathcal W}(\theta)/\partial\theta$ (here we has chosen the
z-axes along $ ({\mathbf M}_{R}\times{\mathbf M}_{L}$)).
 Using this relations one
obtains the equation for the time evolution of the angle $\theta$:
\begin{equation}\label{Theta}
  \frac{M}{g}\frac{d\theta}{dt} = \beta\frac{\partial {\mathcal W}}{\partial\theta}
\end{equation}
On the other hand, multiplying Eq. (1) by ${\mathbf e}_{z}$ and
integrating over the first half-period (0,T), (when the molecular
spin is coupled to the left lead), or over the second one (T,2T),
(when it is coupled to the right lead), we obtain $\Delta m_{z}/T
= g\partial {\\mathcal W}(\theta)\partial\theta$. Combining this
relation with
 Eq. (5)
 one obtainsthe following equation describing time evolution of
the angle $\theta$ :
\begin{equation}\label{evolution}
\frac{1}{T}\Delta m_{z} = (\frac{M}{\beta})\frac{d\theta}{dt}
\end{equation}
The value $\Delta m_{z}/T\equiv\bar{j}$ has a simple physical
interpretation: it gives the average flow of the z-component of
magnetization between the leads, mediated by the periodic
evolution of the mediator magnetization. As a result a mutual
rotation of vectors $\mathbf{M_{\alpha}}$ around x-axis with the
frequency $\Omega = \bar{j}/M$ takes place. To describe the fast
dynamics of {\bf m} it is convenient to use the matrix
representation. Let us introduce the  $(2\times2)$ matrix
$\hat{\rho}$ with the following properties: $Tr\hat{\rho}=0$,
$Tr\hat{\sigma}_{i}\hat{\rho}=2m_{i}/m$, (i=x,y,z and $\sigma_{i}$
are Pauli matrixes). In this case first equation in (1) can be
written in a form:
\begin{equation}\label{rho} \dot{\hat{\rho}} =
-i[\hat{\mathbf{H}}(t),\hat{\rho}] -
\beta[\hat{\rho}[\hat{\rho}\hat{\mathbf{H}}]]
\end{equation}
where
\begin{equation}\label{H}
  \hat{\mathbf{H}}(t)=\frac{1}{2}g
M\bar{J}e^{i\alpha(t)\theta\hat{\sigma}_{z}/4}\hat{\sigma_{x}}e^{-i\alpha(t)\theta
\hat{\sigma}_{z}/4}
\end{equation}
Here we took x-axis in xy-plain along the bisector of the angle
between $\mathbf{M}^{L}$ and $\mathbf{M}^{R}$. Since the
"Hamiltonian" $\hat{\mathbf{H}}(t)$ is a periodic function of
time, the solution of Eq. (6) can be expressed in terms of
"quasienergy" states $|t,\pm\rangle$ defined by the equations:
\begin{eqnarray}\label{states}
i\frac{d}{dt}|t,\pm\rangle =
\hat{\mathbf{H}}(t)|t,\pm\rangle\\
  |t+2NT,\pm\rangle = e^{\pm i\lambda N}|t,\pm\rangle \nonumber
\end{eqnarray}
In this representation the matrix $\hat{\rho}$ has a form:
\begin{eqnarray}\label{rho}
\hat{\rho} = \rho(t)(|t,+\rangle \langle +,t|
- |t,-\rangle \langle -,t|) \nonumber\\
+ \tau(t) |t,+\rangle \langle -,t| + \tau^{\ast}(t) |t,-\rangle
\langle +,t|
\end{eqnarray}
with $\rho^{2} + |\tau|^{2} = 1$. For $\beta = 0$ Eq.(9)  is a
solution of Eq. (6) with $\rho$ and $\tau$ being time independent.
At $\beta \ll 1$ the coefficients $\rho$ and $\tau$ are slow
functions of time. The equation for their time evolution can be
found by substituting of $\hat{\rho}(t)$ in Eq. (6) and averaging
over the period.
 In the case when mediator
 is not coupled to both leads
simultaneously, the states $|t,\pm\rangle$ may be found exactly
and as a result we have the following equation for the $\rho(t)$:
\begin{equation}\label{rho}
  \frac{d\rho}{dt} = \beta g\bar{J}M(1-\rho^{2})C(\theta,\phi)\cos\phi/2
\end{equation}
where $C(\theta,\phi) = (1 -
\sin^{2}\phi/2\cos\theta/2)^{-1/2}\cos\theta/2$.  From this
equation it follows that the molecular spin relaxes to the
periodic regime of evolution ( $|\tau| \rightarrow 0$) and at this
regime $\rho = sign(\cos\phi/2)$.

Now we can calculate $\bar{j} =
m(2T)^{-1}Tr\hat{\sigma}_{z}(\hat{\rho}(T) - \hat{\rho}(0))$.
Making use of the relation (5) we obtain the following equation
for the time evolution of the angle $\theta$:
\begin{equation}\label{F}
  \frac{d\theta}{dt} = -\beta T^{-1}\frac{m}{M}
sign(\cos\phi/2)B(\theta,\phi)\sin\theta
\end{equation}
where $B(\theta,\phi)=\sin^{2}(\phi)/|\sin\lambda|$ From this
equation we can conclude that the relative magnetization of the
leads depends on the $\phi$ -  the angle of precession of
molecular spin during the act of its coupling to the lead. If this
angle corresponds to $(2\pi n,2\pi(n+1))$ the mediated exchange
interaction imposes the ferromagnetic ordering between
single-domain nano-magnets. If $2\phi\in((2n-1)\pi,2n\pi)$ the
angle $\theta$ increases and system demonstrates a trend to
establish the antiferromagnetic ordering. However our analysis
based on the adiabatic approximation breaks for the narrow
interval of angle $\theta$: $|\theta - \pi| \leq m/M \ll 1$. The
angle $\phi = gMJ_{0}e^{A}T$ where $A = V/V_{0}$ is proportional
to the amplitude of alternating electrostatic potential applied to
the tunnel  barriers. Therefore varying the amplitude of
electrical field (or period of oscillation) one can switch
magnetic ordering of the nanomagnets.


To conclude, we suggest a new type of the voltage controlled
exchange coupling between the two nanomagnets when the coupling is
mediated by a small magnetic particle coupled with the nanomagnets
through the tunnel barriers. We demonstrated that the sequentual
periodic suppression of the tunnel barriers with a help of
external electric field allow both ferromagnetic and
antiferromagnetic order in the system. The switch between the two
types of the order can be made by a variation of the parameters of
the controlling ac voltage.
Nanomechanical manipulation of nanomagnets is an alternative to
the above electrical one if ``shuttling of magnetization'' is 
induced by mechanical modulation of tunnel barriers, similarly to
recent experiments \cite{Erbe,Scheible}, where shuttling of electric charge
\cite{GorelikPRL} was observed.

\section{Acknowledgements}
The work is supported by the Swedish Research Council (LYG, RIS)
and by the U.S. Department of Energy, Basic
Energy Sciences-Materials Sciences, under Contract
$\sharp$W-31-109-ENG-38.

\end{multicols}

\end{document}